# Electronic structure of new quaternary superconductors LaONiBi and LaOCuBi from first principles

**I.R. Shein, V.L. Kozhevnikov and A.L. Ivanovskii**

*Institute of Solid State Chemistry, Ural Branch of the Russian Academy of Sciences, Ekaterinburg, GSP-145, 620041, Russia (shein@ihim.uran.ru)*

(Dated: 30 May 2008)

**Based on first-principles FLAPW-GGA calculations, we have investigated the electronic structure of newly synthesized novel superconductors LaONiBi and LaOCuBi, the first bismuth-containing compounds from the family of quaternary oxypnictides which attract now a great deal of interest in search for novel 26-55K superconductors. The band structure, density of states and Fermi surfaces are discussed. Our results indicate that the bonding inside of the (La-O) and (Ni(Cu)-Bi) layers is covalent whereas the bonding between the (La-O) and (Ni(Cu)-Bi) blocks is mostly ionic. For both oxybismuthides, the DOSs at the Fermi level are formed mainly by the states of the (Ni(Cu)-Bi) layers, the corresponding Fermi surfaces have a two-dimensional character and the conduction should be strongly anisotropic andhappen only on the (Ni(Cu)-Bi) layers. As a whole, the new oxybismuthides may be described as *low-$T_C$ superconducting non-magnetic ionic metals*.**

Since the discovery of superconductivity with $T_C \sim 26$K in the fluorine-doped quaternary oxyarsenide LaO$_{1-x}$F$_x$FeAs (x ~ 0.05-0.12),[1] followed by the demonstration [2] that Sr doping in LaOFeAs (La$_{1-x}$Sr$_x$OFeAs, where x ~ 0.09-0.20) leads to a similar value of $T_C$, a new family of layered superconductors (SCs) as a new material platform for further exploration of high-temperature superconductivity has emerged.[3,4] Further promising developments in search for related SCs are achieved with replacing La atoms by other rare-earth metals (*Ln* = Gd,[5] Ce,[6] Sm,[7] Pr and Nd [8,9]), resulting in $T_C$ ~41-55K, i.e. these materials show the highest transition temperatures known except for the high-$T_C$ cuprates. Moreover, comparable values of $T_C$ have been reported for replacing of rare-earth atoms by thorium (Gd$_{1-x}$Th$_x$OFeAs) [10] as well as for non-doped oxygen-deficient samples *Ln*O$_{1-\delta}$FeAs (*Ln* = Sm, Nd, Pr, Ce, La),[11] and the main attention in this area is given at present to a variety of derivatives which are synthesized by replacing *Ln* or oxygen sites in parent phases, namely quaternary oxyarsenides *Ln*OFeAs.

The above oxyarsenides attract now extreme interest because they represent a new group of non-Cu-based layered superconductors adopting relatively high critical temperatures and the upper critical field $H_{c2}$, lie at the border of magnetic instability and should have an unconventional mechanism of superconductivity.[1-11]

On the other hand, these materials belong to a broader family of oxypnictides with a layered 2D-like tetragonal crystal structure (ZrCuSiAs type, space group P4/*nmm*, Z = 2 [12]). For example, a rich set of quaternary 2D oxyphosphides (*Ln*OFeP, *Ln*ORuP (*Ln* = La-Nd, Sm, Gd) and *Ln*OCoP (*Ln* = La-Nd, Sm), [12] as well as LnOOsP, LnOZnP, [13,14] LaONiP, [15,16] LnOMnP,[17] UOCuP,[18] and ThOCuP [19]) has been reported. It is rather interesting that among quaternary oxyphosphides, along with the Fe-containing superconducting phase (LaOFeP, with $T_C$ of 3.2-6.5K [20,21]), the first Ni-based 2D superconductor (LaONiP, with $T_C$ of 3.0-4.3K [15,16]) has been discovered; and later the pure LaONiAs which exhibits bulk superconductivity with $T_C \sim 2.75$ K was found. [22]

Moreover, quite recently *Kozhevnikov et al.* [23] have successfully synthesized a set of new tetragonal oxybismuthides LaO$_{1-\delta}$*M*O (*M* = 3*d* metals). Among them along with a Ni-containing superconducting LaONiBi phase with a critical temperature $T_C$ of 4.2K the first Cu-based 2D superconductor LaOCuBi with $T_C$ of 6K was identified. These oxybismuthides may be considered as possible parent phases for new superconducting materials.

In this Communication we present the results of a systematic study of the electronic properties for the above-mentioned new Ni- and Cu-based superconducting oxybismuthides. These tetragonal phases adopt a layered 2D-like tetragonal crystal structure (ZrCuSiAs type, space group P4/*nmm*, Z = 2 [23]) with a Ni(Cu) atom at 2a (0.75, 0.25, 0), O at 2b (0.75, 0.25, 0.5), La at 2c (0.25, 0.25, $z_{La}$) and Bi at 2c sites (0.25, 0.25, $z_{Bi}$), where the so-called internal coordinates $z_{La}$ and $z_{Bi}$ define the inter-layer distances of La-O and Ni-Bi, respectively, see Ref. [12].

Our band-structure calculations were carried out by means of the full-potential method with mixed basis APW+lo (LAPW) implemented in the WIEN2k suite of programs.[37] The generalized gradient correction (GGA) to exchange-correlation potential in the PBE form [38] was used. La ($5s^2 5p^6 5d^1 6s^2$), O ($2s^2 2p^4$), Ni ($3d^9 4s^1$), Cu ($3d^{10} 4s^1$) and Bi ($5d^{10} 6s^2 6p^3$) are treated as valence states. The calculations were performed with full-lattice optimizations including internal coordinates $z_{La}$ and $z_{Bi}$. The self-consistent calculations were considered to be converged when the difference in the total energy of the crystal did not exceed 0.01 mRy as calculated at consecutive steps. The densities of states (DOSs) are obtained using the modified tetrahedron method. [39]

Two series of calculations were performed: for the nonmagnetic (NM) and magnetic states - in the approximation of FM ordering. For both oxybismuthides, the calculations of the magnetic states are converged into NM states which have the lowest total energies.

The calculated lattice parameters (Table I) are in reasonable agreement with the available experiments; some of the divergences between the theory and experiments should be attributed to oxygen non-stoichiometry of the prepared samples. [23] As can be seen, $a$(LaONiBi) > $a$(LaOCuBi), whereas $c$(LaONiBi) < $c$(LaOCuBi), i.e. *anisotropic deformation* of the crystal structure takes place through replacement of Ni by Cu. Namely, in the sequence LaONiBi → LaOCuBi, the inter-layer distances (La-O)-(Ni,Cu-Bi) increase with simultaneous compression of layers in the *xy* planes. This mechanism can be understood using the data presented in Table I which demonstrate that a substitution of nickel (atomic radius $R^{at}$ = 1.24 Å) by a larger Cu atom ($R^{at}$ = 1.28 Å) results in growth of all bond lengths. At the same time, the Ni-Bi-Ni angles (73.2° for LaONiBi) in LaOCuBi become smaller (64.7°), and this leads to a contraction of Cu-Bi tetrahedra (as compared with Ni-Bi tetrahedra) and a decrease of the cell parameter *a* for the La-Cu oxybismuthide.

In Fig. 1, we show the calculated band structures of LaONiBi and LaOCuBi along some high-symmetry directions of the Brillouin zone (BZ) in the range of energies from -6 eV to +6 eV. The most interesting feature is the 2D-like behavior of the quasi-flat electronic bands along the Γ-Z and A-M directions, which is also found for all other tetragonal oxypnictides. [24-36] We observe that for LaONiBi there are four bands across the Fermi level, whereas for LaOCuBi only three bands intersect the Fermi level owing to increased band filling.

The corresponding Fermi surfaces (FS) in the first BZ are displayed in Fig. 2. In view of the two-dimensional electronic structure, all Fermi surfaces consist of the sheets parallel to the $k_z$ direction. For LaONiBi, these sheets are cylindrical-like, and the first sheet is centered along the *R-X* direction. The other three are centered along the *A-M* high symmetry line. These FS are quite different from the FS of LaOFeAs[24,28]. Though the cylindrical-like Fermi surfaces centered on the *A-M* direction exist for both phases, the tube-like FS centered for LaOFeAs along the Z-Γ direction and the 3D hole pocket around Γ [24,28] for LaONiBi disappear. On the other hand, FSs have similar topologies for isoelectronic LaONiP [36] and LaONiBi.

For LaOCuBi, where just three bands intersect the Fermi energy, and two of them have the crossing point which is located at $E_F$ (Fig. 1), the Fermi surface is made up of three connected sheets. Two of them are cylindrical-like, centered along the *R-X* and *A-M* high symmetry lines (parallel to the $k_z$ direction), while the third sheet is tube-like aligned along the *Z-Γ* direction, Fig. 2.

Figure 3 shows the total DOSs of LaONiBi and LaOCuBi as calculated for the equilibrium geometries. These isostructural materials have some common features of the DOSs. The quasi-core DOSs peaks are located in the following way: peak A: from -21.9 eV to -22.3 eV mainly with Bi 5*d* states; peak B: from -18.6 eV to -20.1 eV mainly with O 2*s* states, peak C: from -14.8 eV to -16.7 eV with overlapping

La 5*p* and O *2s* states, and peak D: from -10.0 eV to -11.3 eV with Bi 6*s* states. The valence band (VB) extends from -5.4 eV to the Fermi level $E_F$ = 0 eV (for LaONiBi) and from -5.2 to $E_F$ (for LaOCuBi). For both oxybismuthides, at the bottom of the conduction band (at about +2.5 eV for LaONiBi and +2.8 for LaOCuBi) there are La *f* states, forming the intensive unoccupied DOS peak E.

Figure 4 shows the total and atomic-resolved *l*-projected valence DOSs of LaONiBi, which extend from -5.4 eV to $E_F$ and include three subbands (A-C). The first subband A ranging from the VB bottom to -3 eV is formed predominantly by O 2*p* states. The next band (B, in the region from -3 eV to -1 eV) contains two intense DOS peaks (Fig. 4) with the main contributions from Ni 3*d* states, together with an admixture from Bi 6*p* states. Finally, the top of the VB (in the interval from -1 eV to $E_F$) is formed basically byNi 3*d* states. This band C (Fig. 1) intersects the Fermi level and continues to +1.5 eV; *i.e.* the near-Fermi region of LaONiBi is formed mainly by the states of the (Ni-Bi) planes.

For LaOCuBi, the increased band filling as a result of the growth of the number of electrons (per cell) leads to a shift of the Fermi level in the region of low binding energies, Fig. 3.

Thus, our calculations give evidence that (i) the electronic bands in the window around the Fermi level are formed mainly by states of (Ni(Cu)-Bi) layers, whereas the bands of (La-O) layers are rather far from the Fermi level; (ii) the general bonding mechanism for both oxybismuthides does not coincide with the "pure" ionic picture and includes covalent interactions inside 2D (La-O) and (Ni(Cu)-Bi) layers, whereas (iii) the bonding between these layers is mostly ionic. The latter can be clearly seen from the 3D electron density distribution for LaONiBi, Fig. 5.

In order to estimate the ionic states of (La-O) and (Ni(Cu)-Bi) layers, we have performed a Bader analysis [40] of the atomic charge density. The obtained results (Table II) show that the charge transfer from (La-O) to (Ni(Cu)-Bi) layers is about 0.54-0.56 e (per formula unit). Thus we can consider the investigated oxybismuthides as *non-magnetic ionic metals*.

Since the near-Fermi electrons are primarily responsible for superconductivity and in order to figure out their nature, the total and orbital decomposed partial DOSs at the Fermi level, $N(E_F)$ are listed in Table III. For both oxybismuthides, the DOSs at the Fermi level are formed by the states of the (Ni(Cu)-Bi) layers, *i.e.* the conduction for these systems is strongly anisotropic and happens only on these layers. In turn, among Ni(Cu) 3*d* states which are responsible for the $N(E_F)$, not only quasi-two-dimensional electronic states derived from Ni(Cu) $d_{xy}$ and Ni(Cu) $d_{x^2-y^2}$ orbitals, but also Ni(Cu) $d_{xz}$, $d_{yz}$, and $d_{z^2}$ orbitals, forming bands with significant dispersion, are present, Table III. The contribution from Bi is noticeable but smaller than that from Ni(Cu) 3*d* orbitals; the main contributions for the $N(E_F)$ are from Bi $6p_{x,y}$ orbitals, increasing considerably as going from LaONiBi to LaONiBi.

These data allow us also to estimate the Sommerfeld constants (γ) and the Pauli paramagnetic susceptibility (χ) for oxybismuthides, assuming the free electron model, as: $\gamma = (\pi^2/3)N(E_F)k_B^2$, and $\chi = \mu_B^2 N(E_F)$. It is seen (Table III) that both γ and χ decrease approximately twice as going from LaONiBi to LaOCuBi; these values also appear appreciably smaller than those obtained for the Fe-containing oxypnictides (from γ = 81 mJ·K$^{-2}$·mol$^{-1}$ for SmO$_{1-x}$F$_x$FeAs [41] to γ = 12.5 mJ·K$^{-2}$·mol$^{-1}$ for LaOFeP [42]), which follows from the differences in the band structures of these species.

In summary, we have systematically studied the band structures, densities of states and Fermi surfaces of new low-$T_C$ superconductors - quaternary oxybismuthides LaONiBi to LaOCuBi, and the bonding picture for these materials has been discussed.

Our results indicate that these phases consist of alternately stacked insulating (La-O) and conductive (Ni(Cu)Bi) layers; the binding between them is mostly ionic. Thus, the newly discovered oxybismuthides may be described as quasi-two-dimensional *non-magnetic ionic metals*, where conduction is strongly anisotropic, happening on the (Ni(Cu)-Bi) layers only.

Finally, the present discussion is focused only on ideal oxybismuthides, whereas numerous issues could be of interest for future studies. For example, in-depth studies are necessary in order to understand the influence of oxygen non-stoichiometry on the properties of these materials. Additionally, these oxybismuthides may be of interest as a new material platform for further exploration of relationships between magnetism and superconductivity. Indeed, in the synthesized today quaternary oxyarsenides-based high-temperature superconductors, an increase in $T_C$ up to ~ 55 K is observed exclusively as a result of electron or hole doping through ion substitutions (or creation of oxygen vacancies) in the Ln-O insulating layers.[1-11] In turn, these high-$T_C$ superconducting phases exist near magnetic instability of oxyarsenide-based SCs, and the magnetic spin fluctuations should play a significant role in unconventional scenarios of superconducting coupling mechanisms which are proposed for these systems. [24-39]

It is possible to assume that for the discussed oxybismuthides the balance between their non-magnetic state and the magnetic instability may be achieved as a result of doping through magnetic 3$d$ ions substitutions (for example, Fe or Co) in conductive (Ni(Cu)Bi) layers. These materials may be very useful as model systems for distinguishing the phonon-mediated low-$T_C$ superconducting mechanism from a situation of an exotic pairing mechanism related to magnetic spin fluctuations.

TABLE I. The optimized lattice parameters ($a$ and $c$, in Å), internal coordinates ($z_{La}$ and $z_{Bi}$) and some interatomic distances ($d$, in Å) for tetragonal oxybismuthides LaONiBi and LaOCuBi.

| phase/parameter | $a$ | $c$ | $z_{La}$ | $z_{Bi}$ |
|---|---|---|---|---|
| LaONiBi | 4.3138 | 8.9619 | 0.1265 | 0.6573 |
| LaOCuBi | 4.1151 | 10.3850 | 0.1168 | 0.6711 |
| phase/parameter | $d$(La-Bi) | $d$(Ni,Cu-Bi) | $d$(Bi-Bi) | $d$(La-Ni) |
| LaONiBi | 3.61 | 2.57 | 4.13 | 3.94 |
| LaOCuBi | 3.65 | 2.72 | 4.59 | 4.48 |

TABLE II. Atomic charges (in e) of quaternary oxybismuthides LaOMBi (where $M$ = Ni and Cu) as obtained from the purely ionic model ($Q^i$), Bader [39] analysis ($Q^B$) and their differences ($\Delta Q = Q^B - Q^i$).

| Atom | $Q^i$ | $Q^B$ | $\Delta Q$ | Atom | $Q^i$ | $Q^B$ | $\Delta Q$ |
|---|---|---|---|---|---|---|---|
| La | 8 (+3) | 9.156 | 1.844 | La | 8 (+3) | 9.146 | 1.854 |
| O | 8 (-2) | 7.303 | -1.302 | O | 8 (-2) | 7.292 | -1.292 |
| Ni | 8 (+2) | 10.130 | 2.130 | Cu | 15 (+2) | 16.994 | 1.994 |
| Bi | 17 (-3) | 15.412 | -1.588 | Bi | 17 (-3) | 15.568 | -1.432 |

TABLE III. Total $N^{tot}(E_F)$ and partial $N^l(E_F)$ densities of states at the Fermi level (in states/eV·atom$^{-1}$), electronic heat capacity $\gamma$ (in mJ·K$^{-2}$·mol$^{-1}$) and molar Pauli paramagnetic susceptibility $\chi$ (in $10^{-4}$ emu/mol) of quaternary oxybismuthides LaOMBi (where $M$ = Ni and Cu).

| Phase/parameter | $N^{M3dxy}(E_F)$ | $N^{M3dx2-y2}(E_F)$ | $N^{M3dxz}(E_F)$ | $N^{M3dyz}(E_F)$ | $N^{M3dz2}(E_F)$ | $N^{Bi6pz}(E_F)$ |
|---|---|---|---|---|---|---|
| LaONiBi | 0.205 | 0.231 | 0.199 | 0.199 | 0.207 | 0.096 |
| LaOCuBi | 0.015 | 0.055 | 0.065 | 0.065 | 0.036 | 0.026 |
| phase/parameter | $N^{Bi6px+y}(E_F)$ | $N^{M3d}(E_F)$ | $N^{Bi6p}(E_F)$ | $N^{tot}(E_F)$ | $\gamma$ | $\chi$ |
| LaONiBi | 0.137 | 0.982 | 0.233 | 2.287 | 5.39 | 0.74 |
| LaOCuBi | 0.172 | 0.236 | 0.198 | 1.037 | 2.44 | 0.34 |

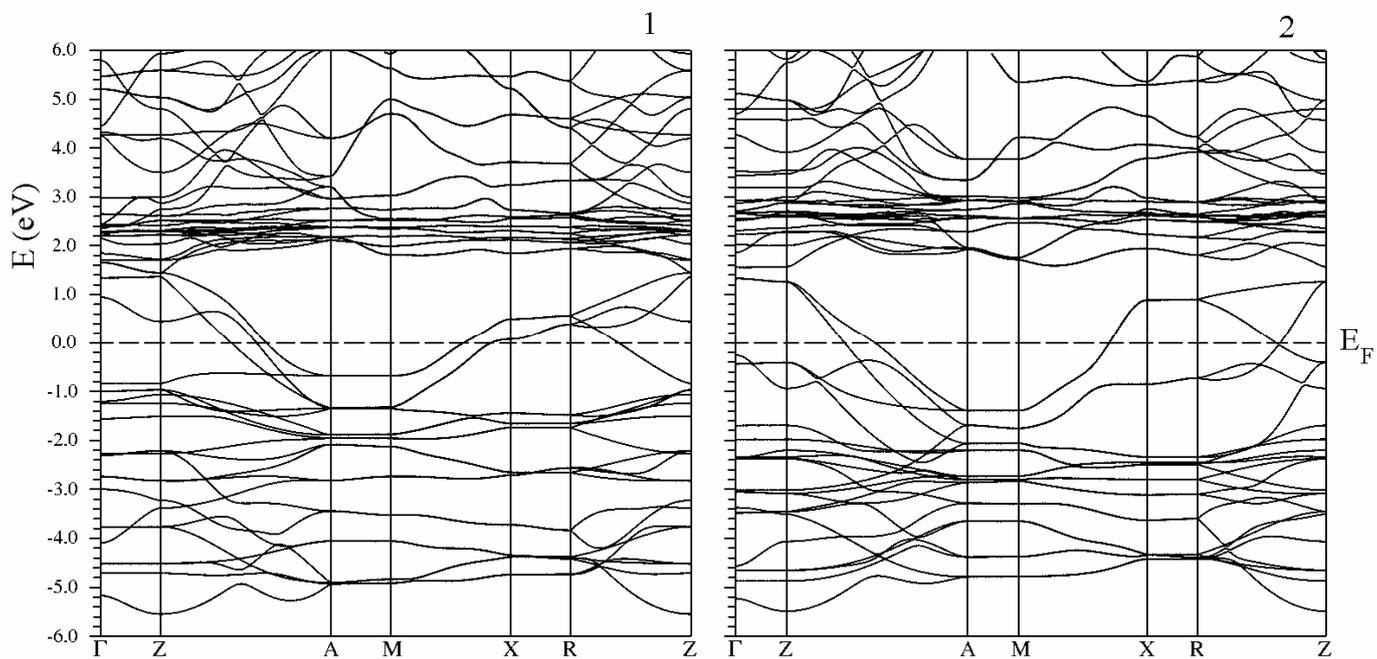

Figure 1. Electronic band structures of LaONiBi (1) and LaOCuBi (2).

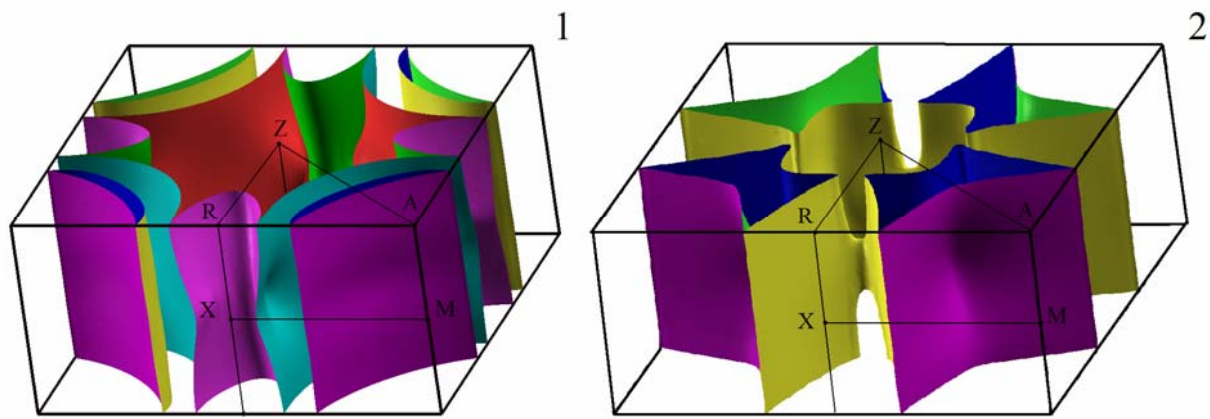

Figure 2. The Fermi surfaces of LaONiBi (1) and LaOCuBi (2).

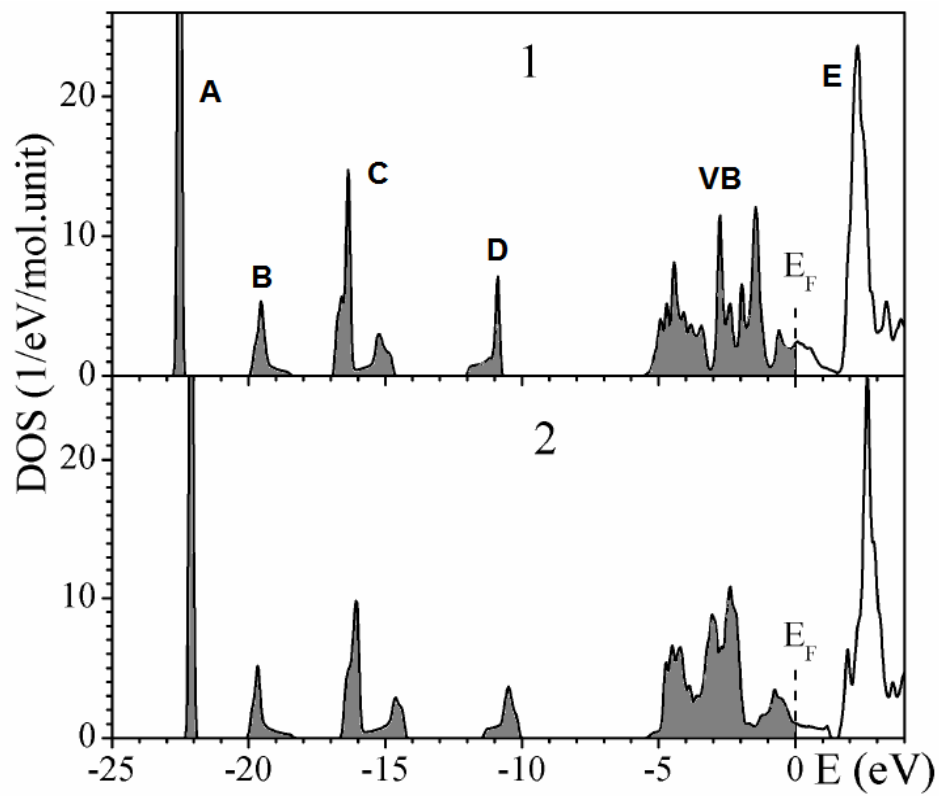

Figure 3. Total densities of states for LaONiBi (1) and LaOCuBi (2).

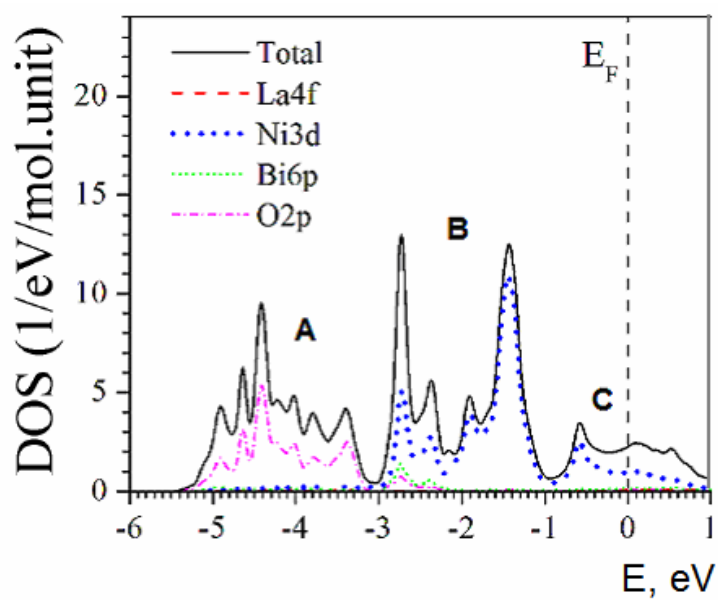

Figure 4. Partial densities of states for LaONiBi.

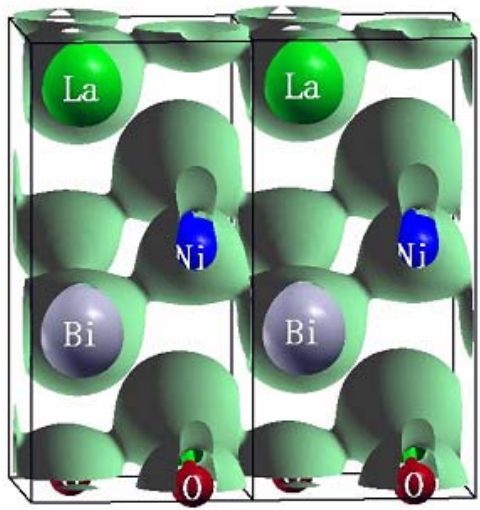

Figure 5. The iso-surface (0.36 $e$/Å$^3$) of charge density for LaONiBi.